\documentclass[letterpaper, 10 pt, conference]{ieeeconf}  

\IEEEoverridecommandlockouts                              

\usepackage{algorithm}
\usepackage{algpseudocode}
\usepackage{kotex}
\usepackage{xcolor}
\usepackage{booktabs}
\usepackage{xspace}
\usepackage{microtype}
\usepackage{booktabs}
\usepackage{siunitx}
\usepackage{multirow}
\usepackage{cite}
\usepackage{amsmath}
\usepackage{amssymb}
\usepackage{amsfonts}
\usepackage[final]{graphicx}
\usepackage{textcomp}
\usepackage{etoolbox}
\usepackage{comment}
\usepackage{hyperref}
\usepackage{cleveref}
\usepackage{flushend}

\title{\LARGE \bf

Do Looks Matter? Exploring Functional and Aesthetic

 Design Preferences for a Robotic Guide Dog
 
}

\author{Aviv L. Cohav$^{*1}$, A. Xinran Gong$^{*1}$, J. Taery Kim$^{1}$, Clint Zeagler$^{1}$, Sehoon Ha$^{1}$, and Bruce N. Walker$^{1}$
\thanks{*co-first authors}
\thanks{$^{1}$Georgia Institute of Technology, Atlanta, GA, USA}
}

\begin{document}

\maketitle
\thispagestyle{empty}
\pagestyle{empty}

\begin{abstract}

Dog guides offer an effective mobility solution for blind or visually impaired (BVI) individuals, but conventional dog guides have limitations including the need for care, potential distractions, societal prejudice, high costs, and limited availability. To address these challenges, we seek to develop a robot dog guide capable of performing the tasks of a conventional dog guide, enhanced with additional features. In this work, we focus on design research to identify functional and aesthetic design concepts to implement into a quadrupedal robot. The aesthetic design remains relevant even for BVI users due to their sensitivity toward societal perceptions and the need for smooth integration into society. We collected data through interviews and surveys to answer specific design questions pertaining to the appearance, texture, features, and method of controlling and communicating with the robot. Our study identified essential and preferred features for a future robot dog guide, which are supported by relevant statistics aligning with each suggestion. These findings will inform the future development of user-centered designs to effectively meet the needs of BVI individuals. 

\end{abstract}


\newcommand{\todo}[1]{\textcolor{blue}{{TODO: #1}}}
\newcommand{\taery}[1]{\textcolor{blue}{{Taery: #1}}}
\newcommand{\sehoon}[1]{\textcolor{red}{{Sehoon: #1}}} 
\newcommand{\wenhao}[1]{\textcolor{blue}{{Wenhao: #1}}} 
\newcommand{\greg}[1]{\textcolor{cyan}{{Greg: #1}}}

\newcommand{\newtext}[1]{#1}
\newcommand{\original}[1]{\textcolor{magenta}{Original: #1}}
\newcommand{\eqnref}[1]{Equation~(\ref{eq:#1})}
\newcommand{\figref}[1]{Figure~\ref{fig:#1}}
\renewcommand{\algref}[1]{Algorithm~\ref{alg:#1}}
\newcommand{\tabref}[1]{Table~\ref{tab:#1}}
\newcommand{\secref}[1]{Section~\ref{sec:#1}}
\newcommand{\mypara}[1]{\noindent\textbf{{#1}.}}

\long\def\ignorethis#1{}

\newcommand{\etal}{{\em{et~al.}\ }}
\newcommand{\eg}{e.g.\ }
\newcommand{\ie}{i.e.\ }

\newcommand{\figtodo}[1]{\framebox[0.8\columnwidth]{\rule{0pt}{1in}#1}}



\newcommand{\pdd}[3]{\ensuremath{\frac{\partial^2{#1}}{\partial{#2}\,\partial{#3}}}}

\newcommand{\mat}[1]{\ensuremath{\mathbf{#1}}}
\newcommand{\set}[1]{\ensuremath{\mathcal{#1}}}

\newcommand{\vc}[1]{\ensuremath{\mathbf{#1}}}
\newcommand{\vEndEff}{\ensuremath{\vc{d}}}
\newcommand{\vRelMove}{\ensuremath{\vc{r}}}
\newcommand{\sSet}{\ensuremath{S}}

\newcommand{\vControl}{\ensuremath{\vc{u}}}
\newcommand{\vPoint}{\ensuremath{\vc{p}}}
\newcommand{\sSpringCoef}{{\ensuremath{k_{s}}}}
\newcommand{\sDamperCoef}{{\ensuremath{k_{d}}}}
\newcommand{\vHandle}{\ensuremath{\vc{h}}}
\newcommand{\vForce}{\ensuremath{\vc{f}}}

\newcommand{\mTransChain}{\ensuremath{\vc{W}}}
\newcommand{\mRotateTrans}{\ensuremath{\vc{R}}}
\newcommand{\sJoint}{\ensuremath{q}}
\newcommand{\vJoint}{\ensuremath{\vc{q}}}
\newcommand{\mJoint}{\ensuremath{\vc{Q}}}
\newcommand{\mMass}{\ensuremath{\vc{M}}}
\newcommand{\sMass}{\ensuremath{{m}}}
\newcommand{\vGravity}{\ensuremath{\vc{g}}}
\newcommand{\vConstr}{\ensuremath{\vc{C}}}
\newcommand{\sConstr}{\ensuremath{C}}
\newcommand{\vCOM}{\ensuremath{\vc{x}}}
\newcommand{\sGeneralForce}[1]{\ensuremath{Q_{#1}}}
\newcommand{\vStateVar}{\ensuremath{\vc{y}}}
\newcommand{\vControlVar}{\ensuremath{\vc{u}}}
\newcommand{\tr}[1]{\ensuremath{\mathrm{tr}\left(#1\right)}}

%
%

\renewcommand{\choose}[2]{\ensuremath{\left(\begin{array}{c} #1 \\ #2 \end{array} \right )}}

\newcommand{\gauss}[3]{\ensuremath{\mathcal{N}(#1 | #2 ; #3)}}

\newcommand{\pctab}{\hspace{0.2in}}
\newenvironment{pseudocode} {\begin{center} \begin{minipage}{\textwidth}
                             \normalsize \vspace{-2\baselineskip} \begin{tabbing}
                             \pctab \= \pctab \= \pctab \= \pctab \=
                             \pctab \= \pctab \= \pctab \= \pctab \= \\}
                            {\end{tabbing} \vspace{-2\baselineskip}
                             \end{minipage} \end{center}}
\newenvironment{items}      {\begin{list}{$\bullet$}
                              {\setlength{\partopsep}{\parskip}
                                \setlength{\parsep}{\parskip}
                                \setlength{\topsep}{0pt}
                                \setlength{\itemsep}{0pt}
                                \settowidth{\labelwidth}{$\bullet$}
                                \setlength{\labelsep}{1ex}
                                \setlength{\leftmargin}{\labelwidth}
                                \addtolength{\leftmargin}{\labelsep}
                                }
                              }
                            {\end{list}}
\newcommand{\newfun}[3]{\noindent\vspace{0pt}\fbox{\begin{minipage}{3.3truein}\vspace{#1}~ {#3}~\vspace{12pt}\end{minipage}}\vspace{#2}}

\newcommand{\key}{\textbf}
\newcommand{\fun}{\textsc}



\section{Introduction}

Professionally trained guide dogs, generically termed `dog guides', play an important role in assisting the lives of individuals who are blind or visually impaired (BVI). Participants have reported that having a dog guide significantly improved their mobility, confidence, and sense of safety when walking or traveling. However, there are certain limitations and concerns associated with owning a live dog guide:
the burden of care and maintenance, potential risk from distractions, and encountering those with an aversion due to allergies, fears, or social prejudice. 
Additionally, the cost of training and maintaining a dog guide is high, with long wait times and limited availability~\cite{wirth2007economic, bender2023aid}. Fewer than 1\% of qualifying adult BVI individuals currently own dog guides~\cite{wirth2007economic}.

Given the limitations of conventional dog guides, we see a need to develop a more accessible and cost-effective solution that will enable more BVI individuals to access the significant benefits a dog guide can provide. In search of a blend of innovation, practicality, and familiarity, we have selected to work on the development of a quadrupedal robot that resembles a dog guide both in shape and function, or a ``robot dog guide''. 
To this end, this study aims to answer design questions on the functional, aesthetic, form factor, and behavioral aspects of a robot dog guide to identify important requirements for future implementations.

\begin{table}[ht]
    \caption{Demographic description of Stage 1 Interview participants}
\vspace{-1em}
    \centering
    \setlength\tabcolsep{3.5pt}
    \begin{tabular}{cccccc}
    \toprule
        \textbf{ID} & \textbf{Gender} & \textbf{Age} & \textbf{Impairment Condition} & \textbf{Mobility Aid}\\ \midrule
        P1 & F & 45-59 & no peripheral vision & Dog Guide\\
        P2 & F & 18-29 & low vision, no night vision & White Cane only\\
        P3 & M & 60-74 & low vision & Dog Guide\\
        P4 & F & 60-74 & low vision & White Cane only\\
        P5 & M & 18-29 & completely blind & White Cane only\\
        P6 & F & 75+ & low vision & Dog Guide\\
        P7 & M & 30-44 & low vision \& hearing loss & Dog Guide\\
        P8 & F & 60-74 & low vision & Dog Guide\\
        P9 & F & 60-74 & low vision \& hearing loss & White cane only\\
        \bottomrule
    \end{tabular}
    \label{tab:bvi-info}
\end{table}

\begin{table}[ht]
\caption{Demographic Description of Stage 2 Survey Participants}
\vspace{-1em}
\centering
\setlength\tabcolsep{3.5pt}
\begin{tabular}{@{} ll *{3}{S[table-format=3.0, table-space-text-post=\%]} @{}}
\toprule
Information&Group &{BVI (\%)} &{ST (\%)}\\
\midrule
Visual impairment&complete/near blindness&50.0&n/a\\
&partially sighted&35.7&n/a\\
&unspecified &14.3&n/a\\
\midrule
Age group&18-24 &31.0&32.8\\
   &25-34 &21.5&53.4\\
   &35-44 &19.0&8.6\\
   &45-54 &9.5&3.4\\
   &55-64&9.5&1.7\\
   &65-74 &9.5&0.0\\
\midrule
Dog guide ownership&DO (current/former owner)&40.5&n/a\\
              &NDO (never owned)&59.5&n/a\\
\bottomrule
\end{tabular}
\vspace{-1em}
\vspace{-1em}
\label{tab:survey-demographics}
\end{table}

While it might seem counter-intuitive to focus on the aesthetics of a robot dog guide for BVI users, research has shown that the visual design of assistive devices can significantly impact social interactions and user experiences. 
Azenkot et al. found that legally blind participants appreciated a navigation robot designed to assimilate into its surroundings, emphasizing that BVI users care about how others perceive their assistive technology. 
The appearance of a robot dog guide can influence the public perception of the user, affecting the user's comfort, confidence, and social acceptance in various environments. 
Moreover, the study highlighted that acceptance from the general public is crucial for the successful deployment of such robots in public spaces. 

To explore these design preferences, we gathered data from BVI participants on their desired appearance, texture, functionalities, and control methods for a robot dog guide, as well as from sighted participants on their impressions of a robot dog guide. 
Participants expressed overall favorable impressions, and the BVI participants highlighted key preferences: the robot should resemble a real dog, appear approachable, and include a clear identifier. They preferred a built-in GPS, Bluetooth connectivity, waterproof materials, and softer textures. They also emphasized multiple control options (voice commands, motion gestures, harness buttons), long battery life, and self-charging capabilities. 
By investigating preferences for aesthetic and form factor design, we seek to create robot dog guides that not only function effectively but also integrate seamlessly into users' lives and social contexts, addressing the concerns of both users and the broader community.
These findings will guide the design of prototypes that meet the specific needs of BVI individuals.

\section{Related Work}
\noindent\textbf{Blind Assistive Technology.}
Prior studies have explored the development of various forms of blind assistive technology. The most studied device has been the technology-assisted white cane, or `smart cane'. Earlier versions of a smart cane were built with ultrasonic sensors to detect objects at various distances, of which the cane would then alert the user through audio output, such as a voice message or beeping tones \cite{wahab2011smartcane, saaid2016canerange}. Later iterations have included more advanced obstacle detection systems, smartphone connectivity, and embedded navigation software, paired with a control panel to allow inputs from the user \cite{subbiah2019caneiot, batterman2018connectedcane, chen2017ccny}.  
However, smart canes, as well as other, less common forms of blind assistive technology, have struggled to be adopted by BVI white cane users, largely due to challenges with design and user interface \cite{kim2013caneusability}. Commonly reported usability issues with smart canes included slow alert times in response to obstacles, difficulty detecting fast-moving objects, inability to detect objects below knee-level (therefore still requiring ground-tapping for floor-level obstacles not unlike with a conventional white cane), and insufficient battery life when in continuous use \cite{kim2013caneusability},\cite{muhammad2010analytical}. 


\noindent\textbf{Robot Dog Guide using Legged Robot.}
Previous work on developing a robot dog guide using legged robots has primarily focused on the control and navigation mechanisms of basic quadrupedal robots.
Developments include leash-guided systems~\cite{xiao2021robotic, morlando2023tethering, chen2023quadruped} and handle-guided systems~\cite{kim2023transforming, hwang2023system}, navigating indoor and irregular terrains. 
They have also begun addressing the complexities of human-robot interactions \cite{kim2023transforming, kim2023train, defazio2023seeing}, developing interaction models and force-responsive controls.
Recent studies have conducted user studies involving BVI participants to examine the potential and design of robot dog guides.
Wang et al. identified usability and trust challenges
in quadrupedal robots compared to wheeled robots~\cite{wang2022can}.
Due analyzed experience of BVI with robot and conventional dog guides, highlighting differences in assistance with navigation and mobility~\cite{due2023guide}.
Hwang et al. studied the handler-dog guide interactions, identifying the need for personalization in robot designs~\cite{hwang2024towards}.
Additionally, Hata et al. tested industrial robots as dog guide, developing user-friendly interfaces such as voice-based apps~\cite{hata2024see}.
These investigations collectively offer insights for designing robot dog guides, emphasizing user experience and effective human-robot communication.


\section{Study Design}
We gathered input from primary stakeholders of the robot dog guide, divided into three subgroups: BVI individuals who have owned a dog guide, BVI individuals who were not dog guide owners, and sighted individuals with generally low degrees of familiarity with dog guides. While the main focus of this study was on the BVI participants, we elected to include survey responses from sighted participants given the importance of social acceptance of the robot by the general public, which could reflect upon the BVI users themselves and affect their interactions with the general population \cite{kayukawa2022perceive}. 

The need-finding processes consisted of two stages. During Stage 1, we conducted in-depth interviews with BVI participants, querying their experiences in using conventional assistive technologies and dog guides. During Stage 2, a large-scale survey was distributed to both BVI and sighted participants. 

This study was approved by the University’s Institutional Review Board (IRB), and all processes were conducted after obtaining the participants' consent.

\subsection{Stage 1: Interviews}
We recruited nine BVI participants (\textbf{Table}~\ref{tab:bvi-info}) for in-depth interviews, which lasted 45-90 minutes for current or former dog guide owners (DO) and 30-60 minutes for participants without dog guides (NDO). Group DO consisted of five participants, while Group NDO consisted of four participants.
All participants were familiar with using white canes as a mobility aid. 

We recruited participants in both groups, DO and NDO, to gather data from those with substantial experience with dog guides, offering potentially more practical insights, and from those without prior experience, providing a perspective that may be less constrained and more open to novel approaches. 

We asked about the participants' overall impressions of a robot dog guide, expectations regarding its potential benefits and challenges compared to a conventional dog guide, their desired methods of giving commands and communicating with the robot dog guide, essential functionalities that the robot dog guide should offer, and their preferences for various aspects of the robot dog guide's form factors. 
For Group DO, we also included questions that asked about the participants' experiences with conventional dog guides. 


\subsection{Stage 2: Large-Scale Surveys} 
After gathering sufficient initial results from the interviews, we created an online survey for distributing to a larger pool of participants. The survey platform used was Qualtrics. 

\subsubsection{Survey Participants}
The survey had 100 participants divided into two primary groups. Group BVI consisted of 42 blind or visually impaired participants, and Group ST consisted of 58 sighted participants. \textbf{Table}~\ref{tab:survey-demographics} shows the demographic information of the survey participants. 

\subsubsection{Question Differentiation} 
Based on their responses to initial qualifying questions, survey participants were sorted into three subgroups: DO, NDO, and ST. Each participant was assigned one of three different versions of the survey. The surveys for BVI participants mirrored the interview categories (overall impressions, communication methods, functionalities, and form factors), but with a more quantitative approach rather than the open-ended questions used in interviews. The DO version included additional questions pertaining to their prior experience with dog guides. The ST version revolved around the participants' prior interactions with and feelings toward dog guides and dogs in general, their thoughts on a robot dog guide, and broad opinions on the aesthetic component of the robot's design. 

\section{Results}

\begin{figure}[t]
\centering
\includegraphics[width=\linewidth]{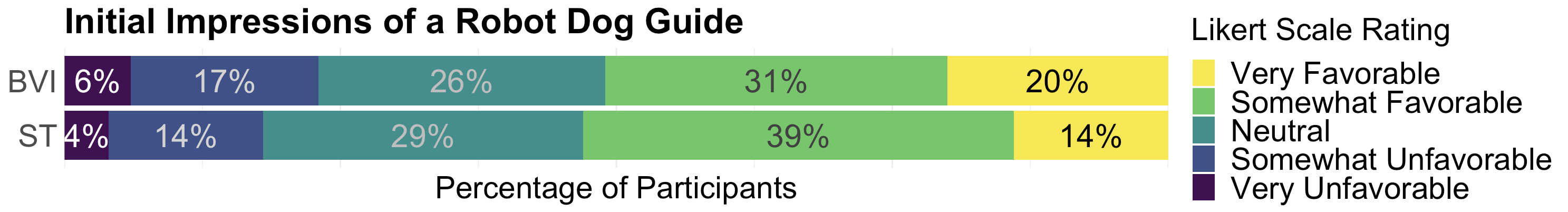}
\vspace{-1em}
\vspace{-1em}
\caption{Participants' initial impressions of the concept of a robot dog guide in Likert scale rating, BVI n = 35, ST n = 56.}
\label{fig:initial_impression}
\vspace{-1em}
\end{figure}

\subsection{Initial Impressions and Concerns}
When asked about their initial reaction to the concept of a `robot dog guide', $51.4\%$ of BVI participants had favorable impressions, $25.7\%$ had neutral impressions, and the remaining $22.9\%$ had unfavorable impressions (\textbf{Fig.}~\ref{fig:initial_impression}).


Both interview and survey participants highlighted several advantages the robot could offer over a conventional dog guide: 1) no requirement for care, training upkeep, and vet costs associated with a living dog; 2) ability of robot to focus on guiding tasks without being distracted by environmental stimuli; 3) potential integration with smart devices and navigation systems. Meanwhile, some concerns raised included: 1) the robot's reliability in ensuring safe navigation (including fear of potential malfunctions); 2) the potential lack of ``intelligent disobedience" or flexibility and adaptability in complex situations; 3) potential need for frequent charging and limited battery life.

\begin{figure*}[bt]
\centering
\includegraphics[height=0.15\textheight]{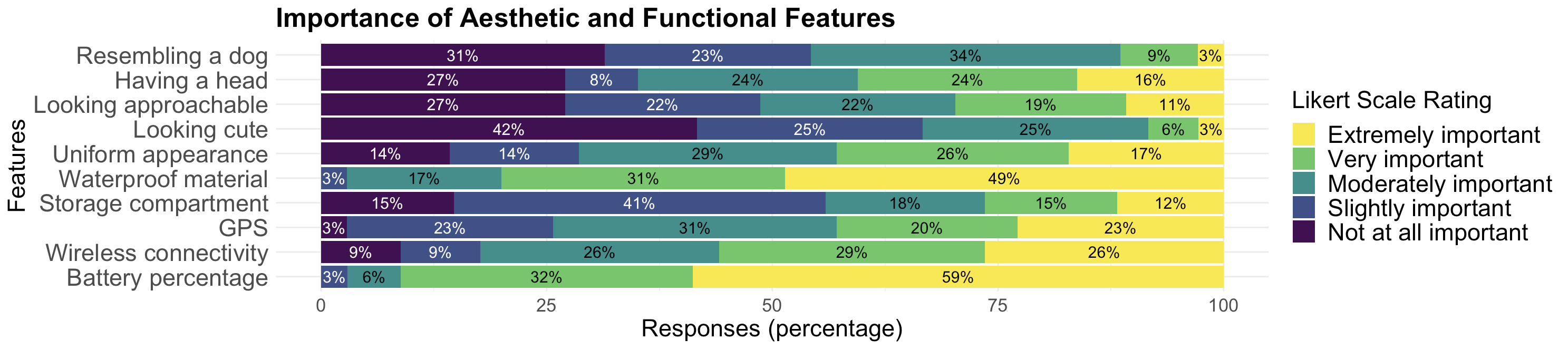}
\vspace{-1em}
\caption{Likert scale ratings from BVI participants on the level of importance of potential features for the robot dog guide to have, n = 37.}
\label{fig:features_importance}
\end{figure*}

\subsection{Form Factors: Appearance and Texture}

\subsubsection{Resemblance to Real Dog}
Interview participants had divided opinions regarding the extent to which the robot dog guide should resemble a real dog, as well as its importance. DO interview participants (P1, P6, P7) generally advocated for a strong resemblance to a real dog with a furry outer surface. They stated that a dog-like appearance would be familiar and comfortable, therefore being more likely to facilitate a smoother transition into society. Meanwhile, other interview participants (P2, P3, P4, P5, P8, P9) believed it would not be necessary for the robot dog guide to resemble a real dog, placing greater emphasis on functional reliability over appearance. For example, P3 and P4 noted that a bipedal robot may perform better than a quadrupedal robot at tasks such as navigating staircases and inclined surfaces. In the survey, 45.7\% of BVI participants showed moderate to strong preference toward a robot that resembles a real dog in appearance, 22.9\% showed a slight preference, and 31.4\% showed no preference (\textbf{Fig.}~\ref{fig:features_importance}). 

\subsubsection{Physical Appeal}
All interview participants indicated that the robot must have a head and some physical appeal, because the appearance of the robot could affect both the robot's and the user's acceptance by society. P6 mentioned that having a head with a nose-like feature could also help the user to correctly identify the robot's orientation through touch. The survey results echoed these findings. A majority of BVI participants found it important for the robot to have a head and appear approachable, with nearly two-thirds also considering it important for the robot to look ``cute" (\textbf{Fig.}~\ref{fig:features_importance}).

\subsubsection{Uniform Identifier}
All interview participants and $85.7\%$ of survey participants indicated that the robot should have a uniform identifier to indicate that it is a working guide (\textbf{Fig.} \ref{fig:features_importance}). Some interview participants (P1, P5, P7) expressed the importance of having a uniform identifier to minimizing unwanted interactions. However, there are differing opinions regarding the specific form of the identifier, such as whether it should be incorporated into the robot dog itself or simply into the harness design, similar to a conventional, white dog guide harness that identifies the dog as a working guide.

\begin{figure}[tbp]
\centering
\includegraphics[width=\linewidth]{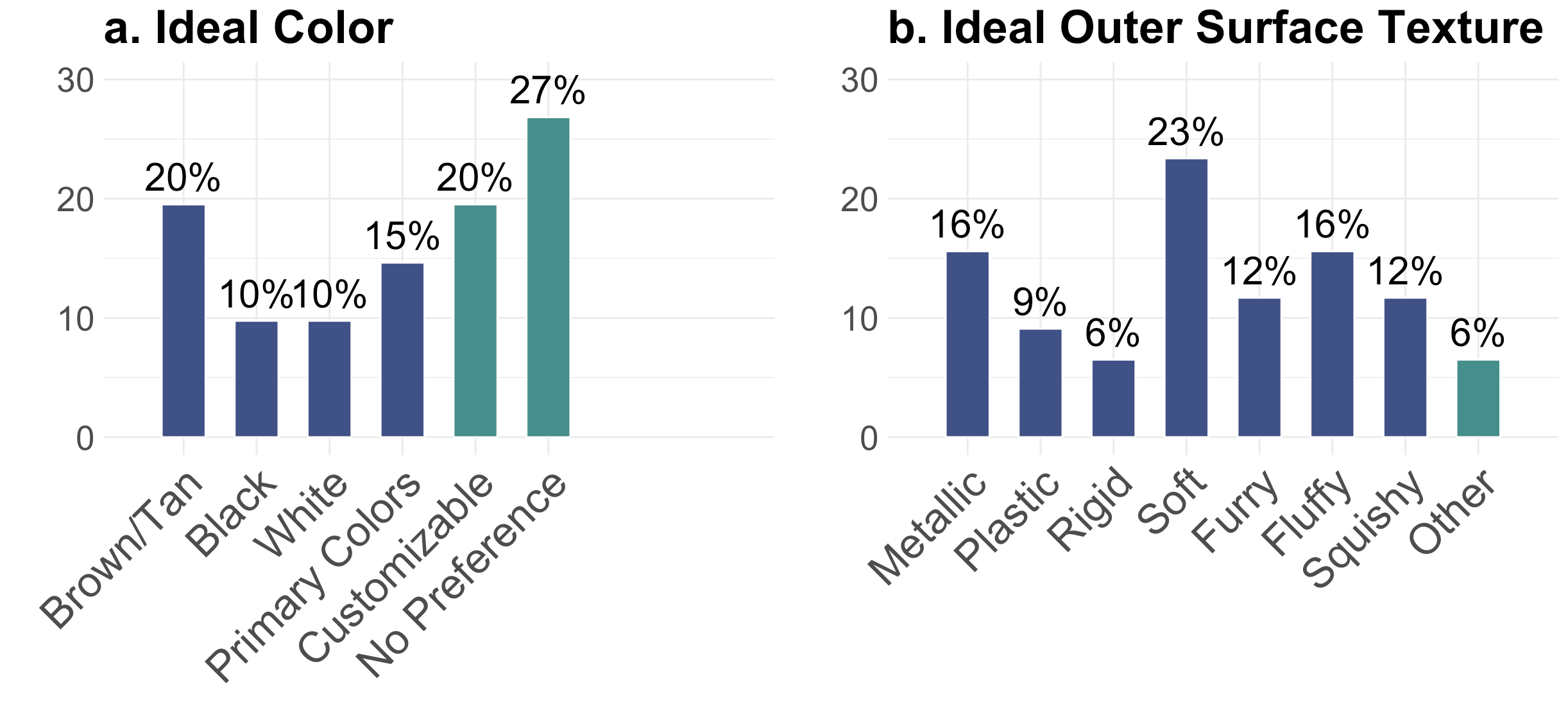}
\vspace{-1em}
\vspace{-1em}
\caption{Distribution of desired color (a) and outer surface texture (b) of robot as indicated by BVI participants, n = 41.}
\label{fig:color_texture}
\end{figure}

\subsubsection{Color}
Interview and survey participants showed no clear preference for the robot's color, favoring customization instead. While 26.8\% had no preference, 19.5\% supported customization, and another 19.5\% preferred a neutral color like a Labrador Retriever (\textbf{Fig.}~\ref{fig:color_texture}a). The rest favored black, white, or a primary/secondary color. A white robot could reflect more light, aiding visibility for partially sighted users and identifying it as a guide.

\subsubsection{Outer Surface, Texture and Material}

Interview participants showed a preference toward the robot dog guide being soft, as having a soft padding would minimize the pain or discomfort caused by inevitable collisions between the robot and the user. P6 specified that even though the outer surface should be soft, the internal structure and body of the robot should be strong enough to sustain impact. P2 and P7 expressed a desire for the robot dog to be ``furry", similar to a live dog. However, there was a consensus that the fabric should also be short, waterproof, and easy to clean. For this reason, P3 and P6 instead expressed a preference for a smooth or rubbery surface. 

As shown in \textbf{Fig.}~\ref{fig:color_texture}b, survey participants preferred a ``soft", ``furry", or ``squishy" texture for the robot over a ``metallic" or ``rigid" one, aligning with interview responses. They also strongly favored the robot being made of waterproof material, with the majority rating it as very or extremely important.

\begin{figure}[tbp]
\centering
\includegraphics[width=\linewidth]{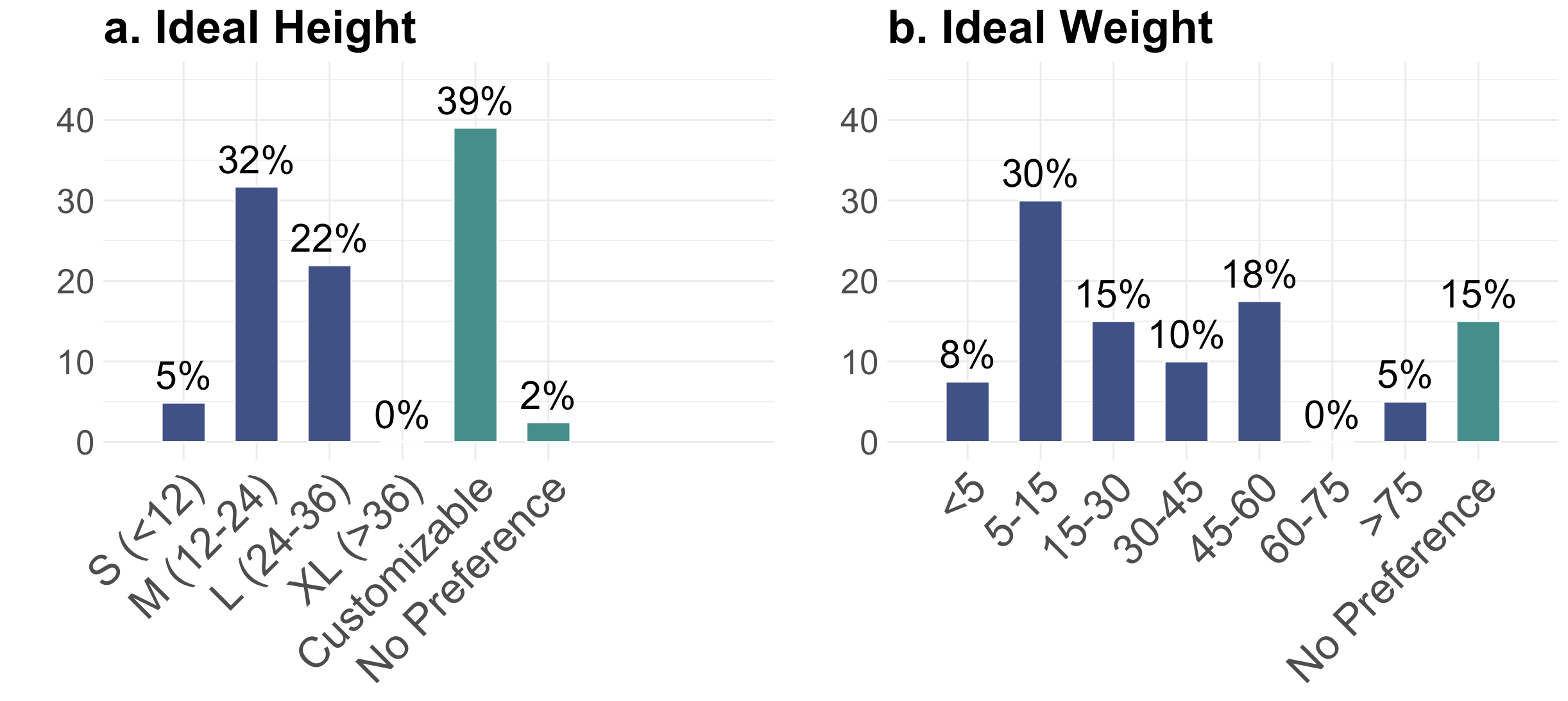}
\vspace{-1em}
\vspace{-1em}
\caption{Distribution of desired height (a) and weight (b) of robot as indicated by BVI participants, n = 41.}
\label{fig:height_weight}
\vspace{-1em}
\end{figure}

\subsubsection{Height and Weight}
From the interviews, P2, P3, P5, and P9 indicated that the robot should resemble a Labrador Retriever or German Shepherd in size. P1, P6, P7, and P8 from group DO favored the robot having an adjustable and/or customizable height. 
They noted the need for adjustable robot legs to accommodate different heights, especially since BVI individuals may hold the harness handle for extended periods and lean on the robot for support.
P1 also suggested customizable pulling force based on the user's strength and body mass.
Although none of the interview participants provided a quantitative number, they showed an overall consensus that the robot should have sufficient mass for users to sense the direction in which they are being led but also be compact (and foldable) and light enough to be lifted for easy transportation.
Survey results showed that most participants preferred the robot to have a customizable height, with significant interest in both `12 to 24 inches' and `24 to 36 inches.' For weight, there was a wide range of preferences, but `5 to 15 pounds' was the most popular choice (\textbf{Fig.} \ref{fig:height_weight}). 


\begin{figure}[ht]
\centering
\includegraphics[width=\linewidth]{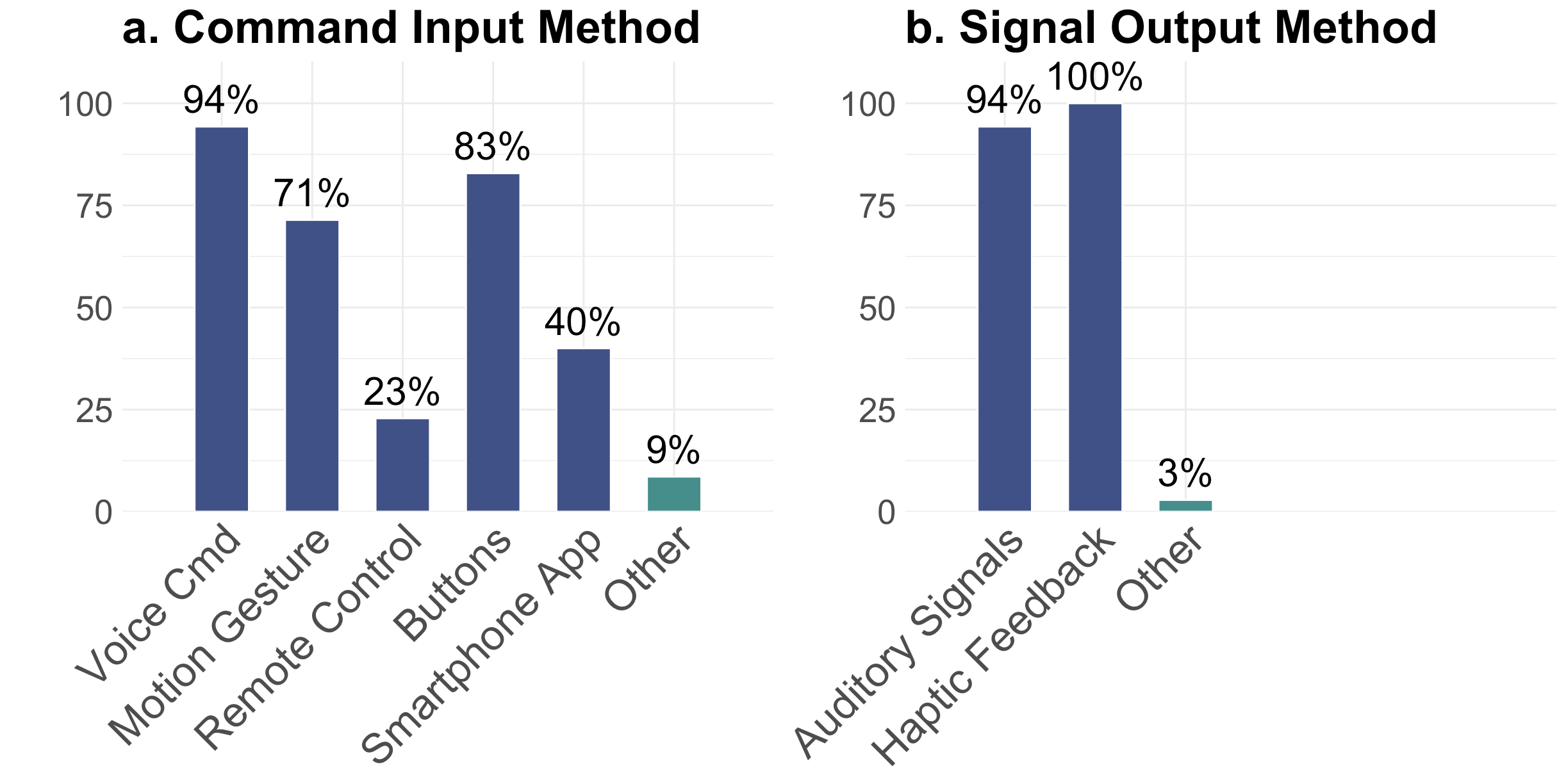}
\vspace{-1em}
\vspace{-1em}
\caption{Desired methods indicated by BVI participants for giving commands to (a) and receiving communication signals from (b) the robot, n = 35.}
\label{fig:communication_methods}
\vspace{-1em}
\end{figure}

\subsection{Commands and Feedback for Control}
All interview participants indicated voice command and auditory feedback as being preferred methods of communication with the robot. They were also in favor of having an added option of voice communication via a Bluetooth earpieces, as well as having alternative command methods such as hand gestures. A suggestion emerged that the robot should have voice recognition technology that distinguishes the owner's voice and responds only to it (as well as potentially a partner or family member's voice when applicable). Several interview participants also showed interest in a simple and comfortable controller pad with support for basic commands, as well as a mobile app for monitoring the robot's maintenance needs. 

Most of the survey participants wanted both voice command and control buttons on the harness handle, as well as some form of motion gestures such as pulling motions of the harness as methods of giving commands to the robot (\textbf{Fig.}~\ref{fig:communication_methods}a). A little under half of the participants also wanted the option of using a smartphone app connected to the robot. A minority of participants wanted remote control as an additional option. All participants wanted haptic feedback (such as vibrations and touch cues), and most also wanted auditory output (\textbf{Fig.}~\ref{fig:communication_methods}b). The ``Other" responses for both questions indicated that it should be customizable, allowing the robot and user to adapt to various situations (e.g., voice command and auditory signals might be ideal in most cases, but button inputs and haptic feedback could be preferred in quiet environments or when it is too loud for effective voice communication). In addition, some users may prefer one method while others may prefer another.

\begin{figure*}[htbp]
\centering
\includegraphics[height=0.09\textheight]{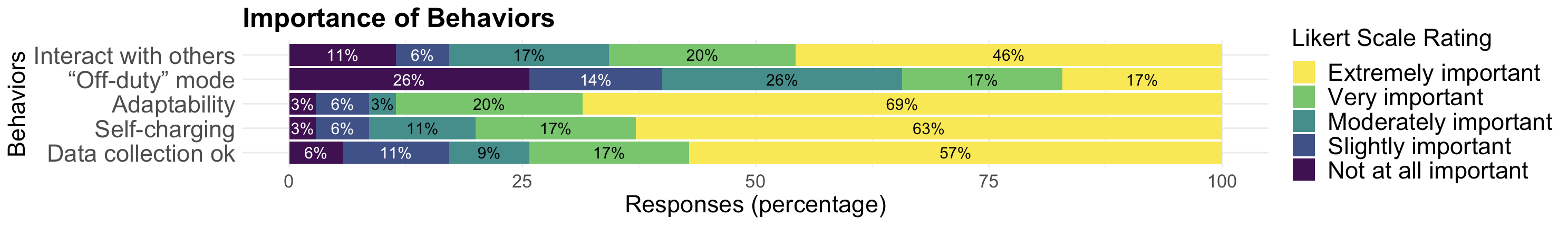}
\vspace{-1em}
\caption{Likert scale ratings from BVI participants on the level of importance of potential behaviors and interactions for the robot guide dog, n = 35.}
\label{fig:behaviors_importance}
\vspace{-1em}
\end{figure*}

\subsection{Behaviors and Social Interactions}
The interview participants had diverging opinions regarding the robot dog guide's social abilities and interactions with others. P1 and P7 indicated that the robot should have an off-duty mode that enables the robot to behave like a pet when safely at home. The participants expressed an appreciation for the warmth and companionship provided by a live dog guide as something they would ``really miss" if they were to have a robot dog guide instead. They were also more open to having the robot interact with and socialize with other people, although it would still be important for the interactions to be limited to only when the off-duty mode is activated. P3, P4, and P5 held neutral opinions, while P2, P6, P8, and P9 held an opposing view. They stated that the robot's functionality and reliability should be the priority, and that adding pet-like behaviors unrelated to the main objective of a mobility aid would not only be unnecessary, but also distracting and potentially even ``creepy''. Specifically, P2 mentioned that she would only want to use the robot as a tool like Google Maps, not for companionship. 
The majority of survey participants felt the robot should interact with other people and pets, although opinions were divided on the desire for an `off-duty' or `pet' mode, with no clear consensus emerging (\textbf{Fig.}~\ref{fig:behaviors_importance}).

We also asked the survey participants to rate the importance of several additional features pertaining to the behaviors and social interactions of the robot (\textbf{Fig.}~\ref{fig:behaviors_importance}). 
On average, the participants completely agreed that the robot should adapt to and learn from the owner’s behaviors and preferences, that it should automatically charge itself, and that the participants would feel comfortable with their data being collected for the purpose of improving service. However, some participants indicated that they are concerned with privacy issues associated with the robot having a camera and preferred to have the option of turning it off.

\subsection{Importance of Aesthetic and Functional Features}
In the survey, we asked BVI participants to rate the importance of several aesthetic and functional features that were mentioned in the interviews (\textbf{Fig.}~\ref{fig:features_importance}). On average, participants rated having a resemblance to a real dog, looking cute, and having a storage compartment as being slightly important. They rated having a head, looking approachable, having a uniform appearance (universal indicator), and having a built-in GPS function as moderately important, having wireless/Bluetooth connectivity as very important, and having a battery percentage notification and the robot being made of waterproof material as being extremely important. 
We performed a T-test for the group difference in the mean ratings of aesthetic factors (having a head, dog resemblance, approachable, cute, uniform) and the mean ratings of functionality factors (GPS, Bluetooth, battery notification, storage compartment, waterproof), and obtained a p-value less than $0.001$. The low p-value suggests that the BVI participants overall placed higher importance on the functionality factors than on the aesthetic factors.

\begin{figure*}[htbp]
\centering
\includegraphics[height=0.16\textheight]{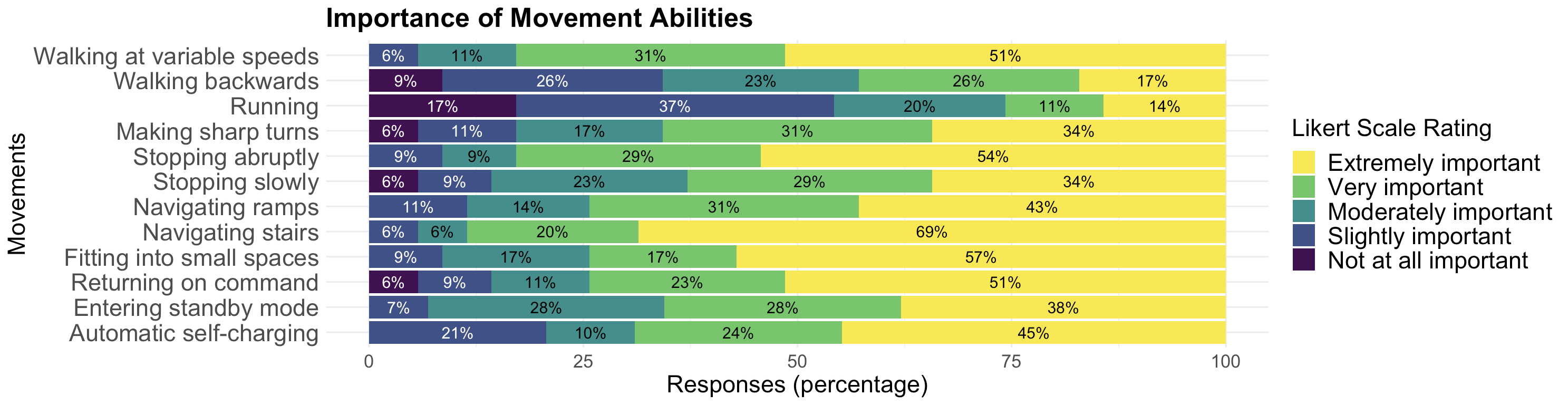}
\vspace{-1em}
\caption{Likert scale ratings from BVI participants on the level of importance of movement abilities that the robot dog guide should have, n = 35.}
\label{fig:movements_importance}
\vspace{-1em}
\end{figure*}

\subsection{Importance of Movement Capabilities}
We asked survey participants to rate the importance of different movement capabilities on a Likert Scale from 1 to 5, with 1 being not important and 5 being extremely important (\textbf{Fig.}~\ref{fig:movements_importance}). On average, the participants rated the robot’s ability to run as being slightly important and the ability to move backwards as being moderately important. All other types of movements asked (variable speeds, sharp turns, abrupt stop, navigating ramps and stairs, crawling into small spaces, returning on command, charging itself, standby) were rated as being very or extremely important. Some other movements not included in the survey but mentioned by participants include: following on command, automatically stopping when dangerous, and navigating in harsh weather conditions. 

\begin{figure}[htbp]
\centering
\includegraphics[width=\linewidth]{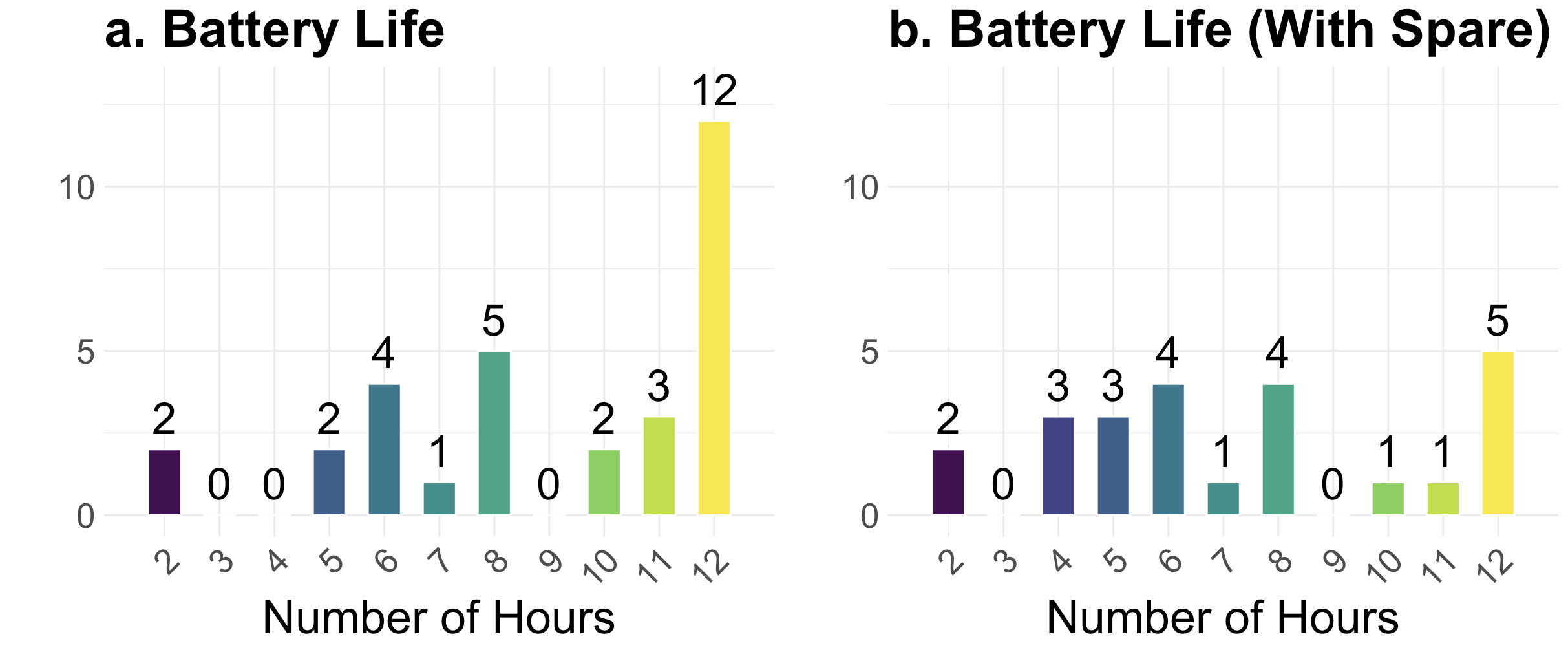}
\vspace{-1em}
\vspace{-1em}
\caption{Minimum number of hours that the battery should last on one charge as indicated by BVI participants for a single battery (a) and if a spare battery is provided (b).}
\label{fig:battery_life}
\vspace{-1em}
\end{figure}

\subsection{Battery}
While the responses varied, the interview participants mostly indicated a need for the battery to last at least 8 to 12 hours with one charge, or to last an entire day. P7 responded favorably to a spare battery being provided. However, P1 and P6 expressed concern that the spare battery would be too heavy to carry around. P6 indicated that the charging process should be done in a wireless manner, similar to the mag-safe charger of an iPhone. 

The majority of the survey participants who responded also indicated that the battery of the robot should last 10 or more hours on one charge (\textbf{Fig.}~\ref{fig:battery_life}). The mean response was 8.84 hours (with 12 as the maximum) with a standard deviation of 3.48 hours. If provided an easily replaceable, spare battery, the participants were overall more amenable to a shorter battery life. In this case, fewer than a third of the participants indicated that the robot should have a minimum battery life of 10 or more hours. The mean response was 7.30 hours, with a standard deviation of 3.26 hours.

\subsection{Responses from Sighted Participants} 
Among ST participants, 21.4\% felt uncomfortable around dogs, and 19.6\% were afraid of them. Of the 24 who expressed fear or dislike of dogs, 37.5\% would feel ``more comfortable with a robot dog than a real dog", though 11.4\% expected they might fear the robot dog. Overall, ST participants had a positive initial impression of a robot dog guide, with 53.6\% reporting favorable views, 28.6\% neutral, and 17.9\% unfavorable—similar to BVI participants (\textbf{Fig.}~\ref{fig:initial_impression}).

ST participants were less supportive of a fur-like texture than BVI participants, with only 17.0\% in favor, while 54.7\% disagreed. However, they placed greater importance on the robot's appearance, with 52.8\% indicating it should look ``cute" and 66.0\% believing it should look ``approachable". Additionally, 77.4\% agreed that a robot dog guide should be allowed everywhere, with few exceptions.



\section{Discussion}

After need-finding and analyzing the collected data, we identified the following requirements for future prototypes of the robot dog guide. 
\subsection{Form Factors}
  \noindent\textbf{Material.} The robot must be made of waterproof material and be easy to clean. It must also have rigid internal structures to sustain reasonable levels of impact with the environment, as well as a soft external padding to prevent injuries to the user from inevitable collisions. 
  
  \noindent\textbf{Height and Weight.} The length of the harness should be customizable to accommodate users with different heights. Extendable legs on the robot itself could also be explored. A conclusive, ideal weight for the robot was not identified, but the robot must have sufficient weight to effectively perform the guiding task, while also being light and compact enough for transportation.
  
  \noindent\textbf{Appearance.} While the robot does not need to exactly resemble a real dog, it needs to have some physical appeal to ensure a smooth transition into society. Having a general shape of a dog with a head is preferred. The robot should have a reasonably approachable presence and not be intimidating. Preferences for color were inconclusive, and many participants favored the option of it being customizable. 
  
  \noindent\textbf{Uniform Identifier.} The robot should have some form of a uniform identifier to identify it as a working guide, either on the robot itself or with a harness that is consistent with the appearance and color of a traditional dog guide harness. 

\subsection{Functionality}
  \noindent\textbf{Navigation.} The robot should have a built-in GPS system and/or be compatible with smartphone navigation apps. 
  
  \noindent\textbf{Battery and Charging.} The charging mechanism should be straightforward, with preference for auto-charging. The battery should last as long as possible, ideally at least 8 hours, with a portable and easily changeable spare battery provided. The robot must notify the user when its battery is low and needs charging, in addition to announcing its current battery percentage when asked. 
  
  \noindent\textbf{Control and Feedback.} The robot must respond to voice commands and recognize essential dog guide commands. Participants also favored having additional control input methods, such as a small, padded controller attached to the harness handle, similar to those on some smart canes. However, comfort should remain a priority when designing the handle. The robot should also be able to provide haptic feedback through handle vibrations, as well as audio feedback both through a speaker and potentially through Bluetooth-connected earphones. In addition, participants showed enthusiasm toward the idea of incorporating an AI interface resembling commonly used AI assistants, which could further enhance usability. 

\subsection{Behavior and Interactions}
  \noindent\textbf{Focus.} The robot must focus solely on the guiding task and not interact with anyone other than the user when working. 
  
  \noindent\textbf{Intelligent Disobedience.} The robot should recognize danger and adapt to complex situations, such as overriding the user's commands when it is unsafe to proceed. The robot should provide the user with a brief explanation afterward. 
  
  \noindent\textbf{Off-Duty Mode.} Some participants were in favor of having an ``off-duty mode''. While having an off-mode is not a requirement, it could be an optional add-on for users who value companionship and affection and wish to interact with the robot like a pet dog when at home.



\section{Conclusion}
In this work, we investigated the preferences of BVI individuals on the aesthetic and functionality design factors of a future robot dog guide through in-depth interviews and a large-scale survey. Our results indicate that the aesthetic factors of the robot desired by the BVI participants include: low-maintenance and waterproof outer surface with soft exterior, uniform identifier, a mechanism to accommodate different heights of users, and some resemblance in appearance to a dog. Functionality requirements of the robot include: built-in or integrated navigation system, easy charging mechanism, battery percentage notifications, and the capacity for multiple methods of communication including voice command, button controls built into the harness handle, auditory signal outputs, and haptic feedback. We also gathered feedback on desirable behavior and interaction patterns, which include: the ability to remain focused on the guiding task, displaying intelligent disobedience when necessary, and an ``off-duty'' mode as a potential option. 

For future work, it may be worthwhile to test physical prototypes of different heights and weights of the robot, given that the participants often had some difficulty stating a specific desired height and weight without a physical reference point. We also plan on further refining the specific texture of the robot's outer surface, the degree of resemblance to a dog, and the specific functions of the individual buttons built into the harness handle through further design studies with participants. Additionally, we are interested in developing a voice user interface (VUI)-enabled AI agent for the robot to maximize a comfortable and supportive user experience. 



\section*{ACKNOWLEDGMENT}
This research has been funded by the Industrial Technology Innovation Program (P0028404, development of a product-level humanoid mobile robot for medical assistance equipped with bidirectional customizable human-robot interaction, autonomous semantic navigation, and dual-arm complex manipulation capabilities using large-scale artificial intelligence models) of the Ministry of Industry, Trade and Energy of Korea.
JK was supported by the NSF GRFP under Grant No. DGE-2039655. Any opinion, findings, and conclusions or recommendations expressed in this material are those of the author(s) and do not necessarily reflect the views of the National Science Foundation, or any sponsor.

\bibliographystyle{IEEEtran}
\bibliography{bib}


\addtolength{\textheight}{-12cm}   


\end{document}